\documentclass[final,3p,times]{elsarticle}

\usepackage{lipsum}
\usepackage{graphicx}
\usepackage{amsthm} 
\usepackage{amsmath,amssymb,physics}
\usepackage{hyperref}
\usepackage{url} 
\usepackage{lineno} 
\usepackage{xspace}
\usepackage{braket}
\usepackage{wrapfig}
\usepackage{courier}
\usepackage{xcolor,ulem} 
\usepackage{comment} 
\usepackage{array} 
\usepackage{
soul}
\usepackage{longtable}
\usepackage{diagbox}
\usepackage{xparse}
\usepackage{subfigure}
\usepackage{listings} 
\usepackage{rotating}

 \usepackage{setspace}

\definecolor{codegreen}{rgb}{0,0.6,0}
\definecolor{codegray}{rgb}{0.5,0.5,0.5}
\definecolor{codepurple}{rgb}{0.58,0,0.82}
\definecolor{backcolour}{rgb}{0.95,0.95,0.92}

\definecolor{nicegreen}{rgb}{0.,0.5,0.}

\definecolor{hotpink}{rgb}{1,0.27.0.635}

\begin{document}

\begin{frontmatter}

\begin{flushright}
Preprint MSUHEP-26-008
\end{flushright}

\title{Probing SU(2) Quark Flavor Asymmetry with $W$ Bosons at RHIC}

\author[one]{\mbox{Maximiliano Ponce-Chavez\fnref{fnMPC}}}

\affiliation[one]{organization={Department of Physics and Astronomy,
Michigan State University}, 
            city={East Lansing},
            postcode={48824}, 
            state={MI},
            country={USA}}

\begin{abstract}
This work examines the effects of multiple parton radiation on proton-proton $W^\pm$ boson production at the Relativistic Heavy Ion Collider (RHIC), focusing on the ratio of charged lepton pseudorapidity differential cross sections to test the SU(2) flavor symmetry violation in the proton quark sea. I show that the measured ratio is marginally affected by the fiducial constraints imposed on the QCD recoil radiation by the STAR experiment, and that the ratio is also stable with respect to transverse momentum resummation effects. Hence, the STAR data on the $W^+/W^-$ cross section ratio provides a robust discrimination of $\bar{d}(x)$ and $\bar{u}(x)$ parton distribution functions (PDFs) at momentum fractions of order 0.1, shown by comparing predictions with various PDF models and using the $L_2$ sensitivity analysis and reweighting methods.

\end{abstract}

\end{frontmatter}

\section{Introduction}

Parton distribution functions are fundamental objects that parameterize the internal structure of hadrons in terms of probabilities dependent on the fractional momentum ($x$) carried by its constituents, namely quarks and gluons, at various resolution energy scales. PDFs are essential for making predictions for hard-scattering experiments involving hadrons, whether to perform tests to the Standard Model or to search for signatures of new physics. Due to their universality (independence from the hard-scattering process \cite{CTEQ:1993hwr}), phenomenological parameterizations can be obtained by performing global fits of experimental measurements which are sensitive to PDFs, with the achieved precision depending on both theoretical and experimental inputs, such as the perturbative order of the Quantum Chromodynamics (QCD) elements (scattering amplitudes, splitting functions, etc.), sensitivity of individual experiments to specific PDF flavors \cite{Jing:2023isu}, and statistical/systematic uncertainties of the experimental data going into the fit. In particular, deep inelastic scattering (DIS) and Drell-Yan pair production experiments are key for separating the proton's valence and sea components, although the latter remains to be well-charted, since the underlying mechanism countering an early parton model assumption of SU(3) flavor symmetry ($\bar{u}(x) = \bar{d}(x) = \bar{s}(x)$) is still under exploration. A summary of past scattering experiments reporting non-negligible deviations from 1/3 in the Gottfried sum rule \footnote{Gottfried Sum Rule: $\int_{0}^1 dx \frac{1}{x} [F_2^p(x) - F_2^n(x)] = \frac{1}{3} + \frac{2}{3}\int_0^1 dx [\bar{u}^p(x) - \bar{d}^p(x)]$. Values different from 1/3 imply $\bar{d}(x) \neq \bar{u}(x)$.}, signaling sea quark flavor asymmetry, is found in Ref. \cite{Peng:1999es}. Dedicated experiments, such as NMC \cite{NewMuon:1991hlj,NewMuon:1993oys}, NuSea \cite{NuSea:2001idv}, and SeaQuest \cite{SeaQuest:2021zxb,FNALE906SeaQuest:2025kjo}, measured the ratio of proton-deuteron and proton-proton cross sections, effectively constraining $\bar{d}(x)/\bar{u}(x)$ at moderate and high $x$, establishing thereupon that the sea-flavor asymmetry is $x$-dependent. Precise determination of such dependence remains remains an open question, as measurements from the NuSea and SeaQuest experiments at Fermilab display some tension \cite{SeaQuest:2021zxb}.\\

Complementing fixed-target measurements, collider-based experiments, \textit{e.g.} the Large Hadron Collider (LHC) \cite{CMS:2011bet,ATLAS:2025ede} and Tevatron \cite{D0:2014kma,CDF:2021kyj}, place constraints on the proton quark sea through inclusive measurements of lepton charge asymmetry. Furthermore, proton-(anti)proton interactions in collider settings are free from nuclear effects, unlike fixed-target experiments where such effects are assumed to be minimal, but may still impact $\bar{d}(x)/\bar{u}(x)$, pointing to the question of whether the observed asymmetry truly stems from differing contributions in the proton quark sea.\\

This work focuses on the STAR experiment at BNL's Relativistic Heavy Ion Collider, which explores the flavor asymmetry via the $W^\pm$ cross section ratio of proton-proton scattering at transfer momentum scales significantly larger than $\Lambda_\textrm{QCD}$. The STAR collaboration has performed these measurements in sequential runs with unpolarized proton beams; 2011, 2012, 2013 and 2017, with the first three combined in a single dataset \cite{STAR:2020vuq}, while the latest was recently published in Ref. \cite{nam2022measurementswzgammacross}. The ratio of $e^\pm$ pseudorapidity differential cross sections in inclusive production, $pp\to (W^\pm\to e^\pm\nu) X$, is sensitive to $\bar{d}(x,Q) /\bar{u}(x,Q)$ at an energy scale of order of the boson's virtuality $Q \approx M_W \approx80.4$ GeV. At STAR, $W^\pm$ decay positrons and electrons are registered in the pseudorapidity fiducial region $-1<\eta<2$, yielding an approximate fractional momentum coverage of $0.06 < x < 0.4$. The event selection procedure imposes vetoes on charged leptons and jets to reject additional sources of $e^\pm$ production and suppress the large QCD background; these constraints on QCD radiation generally break down the inclusivity of the lepton production process required to apply perturbative QCD factorization. Therefore, the fixed-order $\sigma_{W^+}/\sigma_{W^-}$ ratio calculations in Ref. \cite{STAR:2020vuq} may be potentially modified by higher-order terms emerging due to the non-inclusive veto on the QCD recoil. This article quantifies the significance of these potential corrections by categorizing radiative contributions arising in the $\sigma_{W^+}/\sigma_{W^-}$ ratio and performing all-order resummation of parton radiation at two perturbative orders.\\

First, I compare theoretical predictions at the fixed next-to-leading (NLO) and next-to-next-to-leading (NNLO) orders in the strong force coupling ($\alpha_s$) to those arising from the Collins, Soper and Sterman transverse momentum resummation formalism \cite{Collins:1984kg}, to evaluate the effects of collinear and soft QCD emissions on the observable $e^\pm$ pseudorapidities. Second, I compare resummed predictions inclusive in QCD recoil to those from the \textsc{Pythia} parton showering code \cite{Bierlich:2022pfr}, in which the STAR-like veto on QCD recoil is imposed, to quantify to what extent the veto leads to deviations. Finally, I compare cross section ratio measurements to analyze the power of STAR data to discriminate among contemporary PDF sets: CT18 \cite{Hou:2019efy}, MSHT20 \cite{Bailey:2020ooq}, the Hessian versions of NNPDF3.1 \cite{NNPDF:2017mvq} and 4.0 \cite{NNPDF:2021njg}, and JAM24 \cite{Anderson:2024evk} for fixed-order NLO predictions. The resummed differential distributions are computed with the \textsc{ResBos}/\textsc{ResBos2} \cite{Balazs:1997xd,Isaacson:2017hgb,Isaacson:2023iui,Isaacson:2022rts} code, up to next-to-next-to-next-to-leading logarithm (N$^3$LL) matched to perturbative NNLO. Fixed-order distributions are computed with the \textsc{MCFM-10.3} code \cite{Campbell:1999ah,Campbell:2011bn,Campbell:2019dru}.\\

This paper is organized as follows: Section \ref{sec:ResFO} briefly summarizes the differences between vector boson production with the fixed-perturbative order and resummation prescriptions. Section \ref{sec:PYTHIA} reviews the differences in the assumptions from STAR's \textsc{Pythia} simulations shaping the experimentally extracted signal and those from the resummed approach. The comparison between experimental data and theoretical predictions with different choices of perturbative orders and PDFs is provided in Section \ref{sec:Data}. Section \ref{sec:Sens} details the impact of the new set of measurements mainly on the $\bar{d}(x,Q)$ and $\bar{u}(x,Q)$ distributions of the CT18 PDF set through the $L_2$ sensitivity study, complemented by a reweighting analysis done on a new PDF set from the CTEQ-TEA collaboration. Additional observations and concluding remarks are presented in Section \ref{sec:Conc}.

\section{Fixed-order and resummed computations for inclusive boson production}
\label{sec:ResFO}
\begin{figure}[!ht]
\centering
\includegraphics[width=0.8\textwidth]{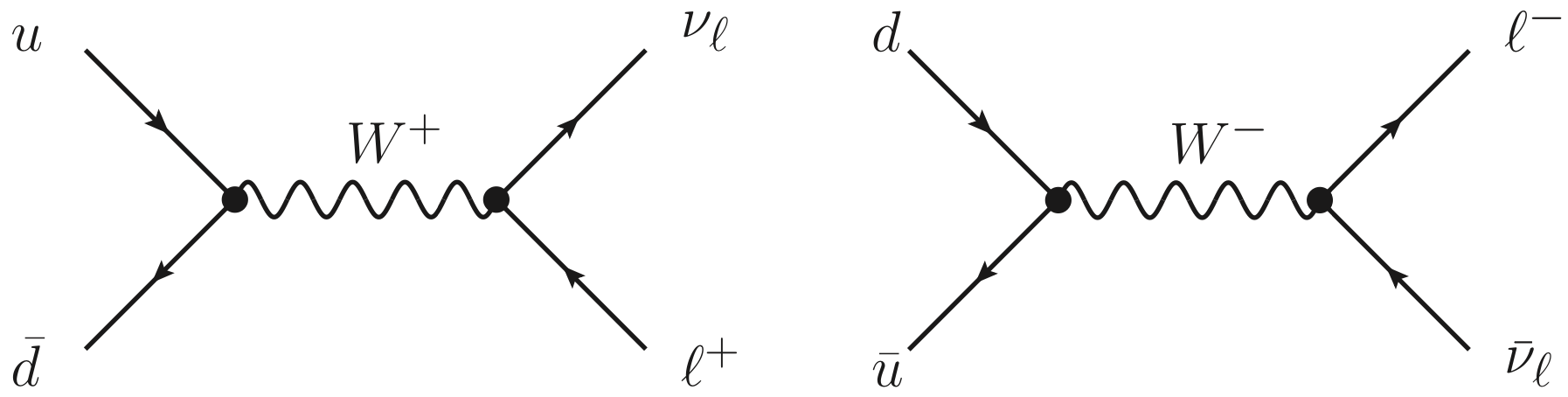}
\caption{Leading Born-level charged-current vector boson production \cite{Campbell:2017hsr}.}
\label{fig:Feyn}
\end{figure}

The Drell-Yan (DY) process is central to exploring hadronic structure and testing the validity of perturbative QCD, while also being a high-precision probe for measurements of the Standard Model electroweak parameters. The DY hadronic cross section is given by the convolution of a partonic cross section $\hat{\sigma}$, computed as a perturbative expansion in $\alpha_s$, with PDFs $f_{i/h}(\xi,\mu_F)$ at the characteristic energy scale $Q$ of the partonic process,
\begin{equation}
    \sigma_{h_1 h_2 \to l \nu_l} = \sum_{i,j=q,\bar{q},g} \int d\xi_1 d\xi_2 \left[ f_{i/h_1}(\xi_1,Q)\ f_{{j/h_2}}(\xi_2,Q) \right] \hat{\sigma}(\xi_1,\xi_2),
\end{equation}
where the sum runs over all parton flavors within the hadron. In the charged current case at the Born level, up and down-type quarks and antiquarks from various generations annihilate to produce $W$ bosons with null transverse momentum ($q_T=0$), further decaying into a lepton and the associated neutrino as shown in Fig. \ref{fig:Feyn}. Omitting the PDF scale dependence, the proton-proton differential cross section ratio at the Born level in terms of charged boson rapidity $y_{W^\pm}$ is:
\begin{equation}
\begin{split}
    \frac{d\sigma^{W^+}/dy_{W^+}}{d\sigma^{W^-}/dy_{W^-}} &\varpropto \frac{ |V_{ud}|^2 \left[u(\xi_1) \bar{d}(\xi_2) + \bar{d}(\xi_1) u(\xi_2)\right] + |V_{us}|^2 \left[u(\xi_1) \bar{s}(\xi_2) + \bar{s}(\xi_1) u(\xi_2)\right] + \ldots}{ |V_{ud}|^2 \left[\bar{u}(\xi_1)d(\xi_2) + d(\xi_1)\bar{u}(\xi_2)\right] + |V_{us}|^2 \left[\bar{u}(\xi_1)s(\xi_2) + s(\xi_1)\bar{u}(\xi_2)\right] + \ldots} \\
    &\underset{\text{Leading}}{\approx} \frac{ u(\xi_1) \bar{d}(\xi_2) + \bar{d}(\xi_1) u(\xi_2) }{ \bar{u}(\xi_1)d(\xi_2) + d(\xi_1)\bar{u}(\xi_2) },
\end{split}
\label{eq:rat}
\end{equation}
with $V_{ij}$ the elements of the Cabibbo-Kobayashi-Maskawa quark flavor mixing matrix, and fractional momenta satisfying $\xi_1 \xi_2 = Q/\sqrt{s}$, where $\sqrt{s}$ is the center-of-mass energy. Within the kinematic coverage of the STAR experiment, Eq. (\ref{eq:rat}) is approximately equal to the leading term in the second line due to the suppression of cross-generation flavor mixing and heavy-quark contributions. In the STAR kinematic region, $u(x,Q)$ and $d(x,Q)$ PDFs are well constrained by DIS data so, in principle, a measurement of the ratio in Eq.~(\ref{eq:rat}) would allow the extraction of information about $\bar{d}/\bar{u}$. However, a direct measurement is not feasible because undetectable neutrinos in the final state impede a unique determination of the $W^\pm$ boson rapidities. Instead, the STAR collaboration measures the cross section ratio as a function of the charged lepton pseudorapidity $\eta_{e^\pm}$, which is closely related to the boson rapidity \cite{Nadolsky:2003fz}. Moreover, the STAR experiment imposes vetoes on hadronic recoil against the candidate charged leptons, further addressed in Section~\ref{sec:PYTHIA}. 

\begin{figure}[!ht]
\centering
\includegraphics[width=0.495\textwidth]{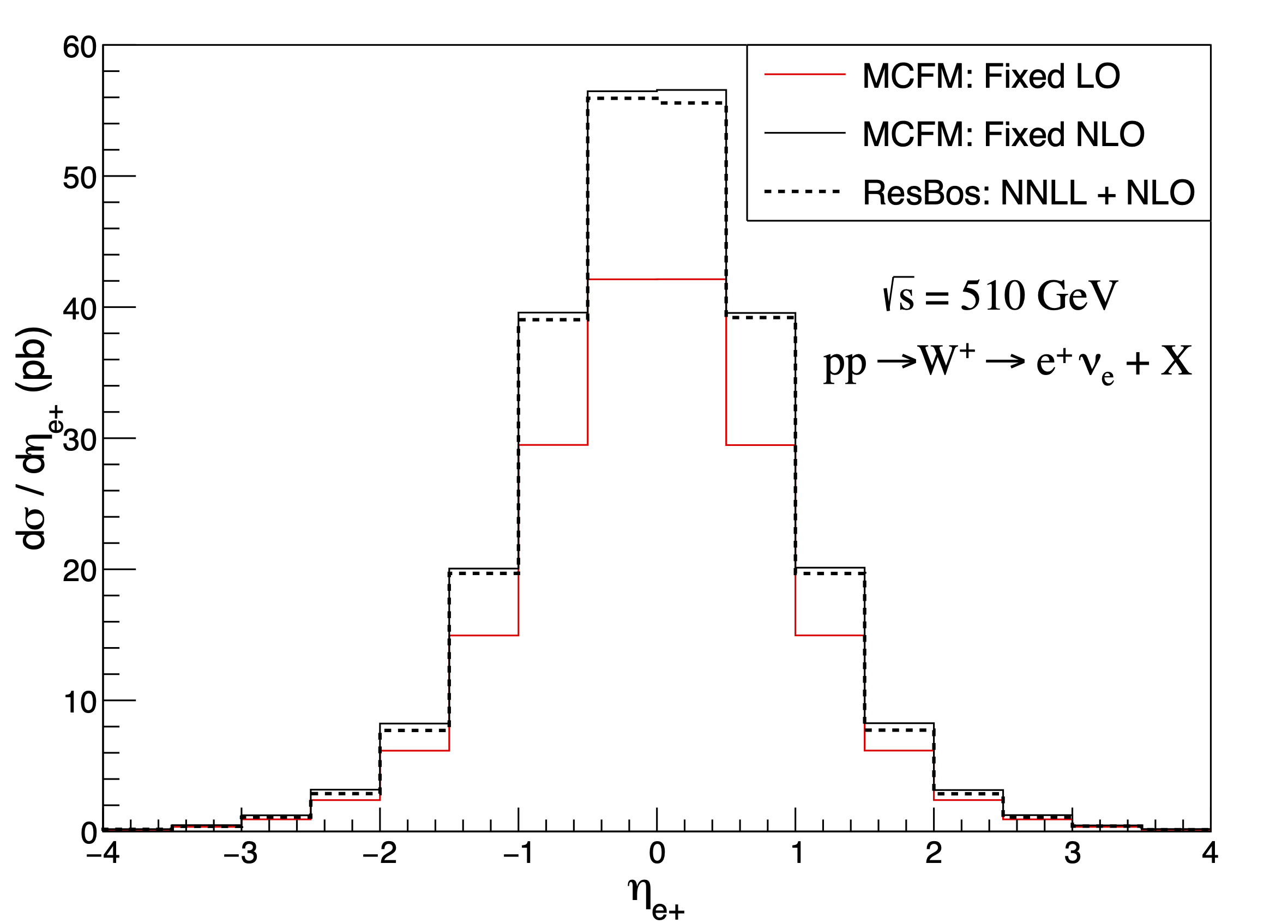}
\includegraphics[width=0.495\textwidth]{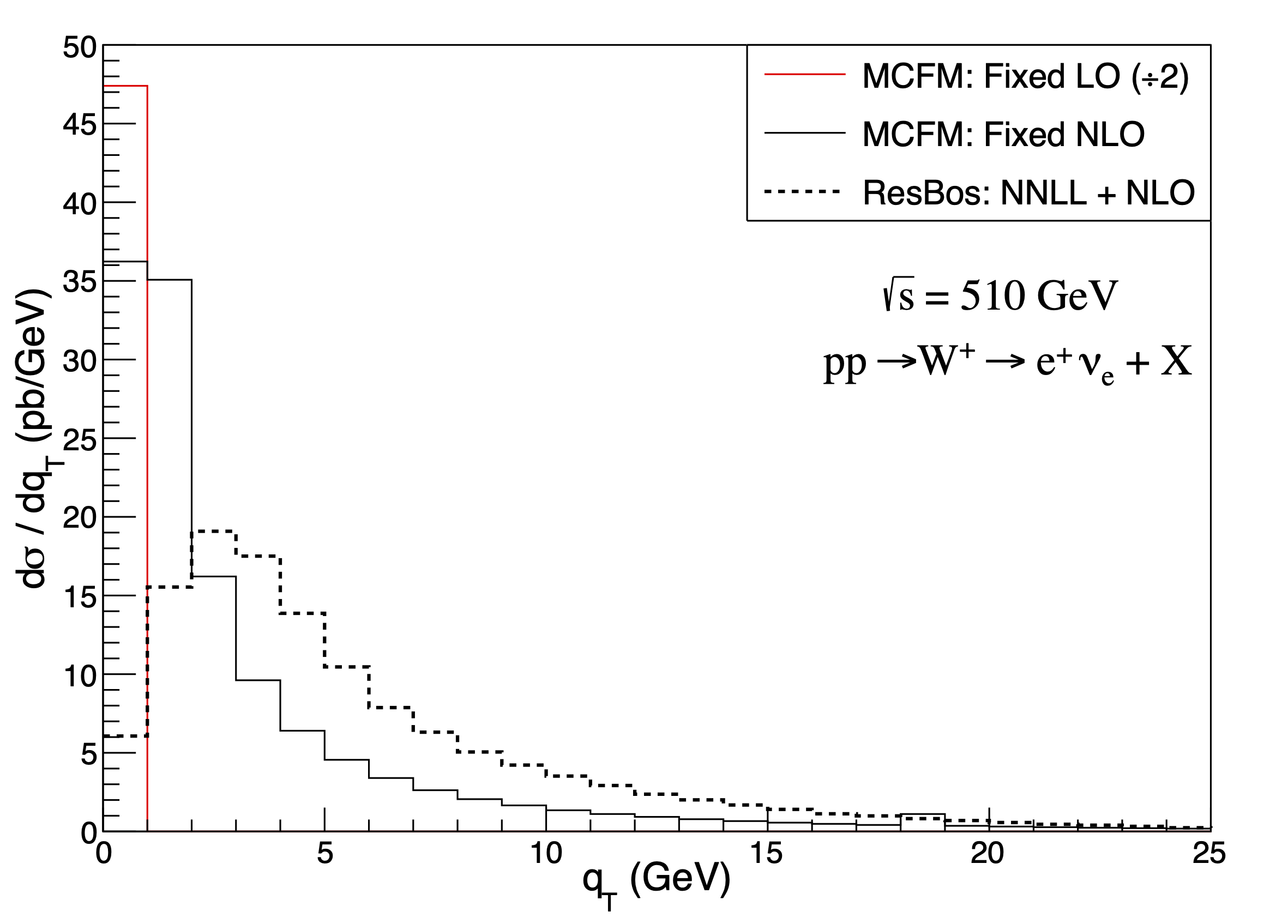}

\caption{Positron pseudorapidity (left) and vector boson transverse momentum (right) differential cross sections for inclusive $W^+$ production at $\sqrt{s}=510$ GeV with fixed-order (MCFM) and resummed (ResBos) calculations at LO and NLO, using the CT18 NNLO PDF set.
} 
\label{fig:dists}
\end{figure}

Fig.~\ref{fig:dists} shows lepton pseudorapidity and $W$ boson transverse momentum differential cross sections that are inclusive in QCD radiation, computed up to the fixed NLO, as well as resummed next-to-next-to-leading logarithm matched to NLO. Radiative contributions notably enhance the normalization of the $W$ production rate by a factor $K_\textrm{NLO} \equiv \sigma_\textrm{NLO}/\sigma_\textrm{LO}\approx 1+\frac{\alpha_s}{2\pi}C_F(1+\frac{4\pi^2}{3})$ visible in the left panel. As the dominant part of $K_\textrm{NLO}$ arises from the hard part of the virtual correction, it is about the same for the fixed-order NLO and resummed cross sections, depending weakly on the rapidity and charge of the boson.\\

\begin{figure}[p]
\centering
\includegraphics[width=0.85\textwidth]{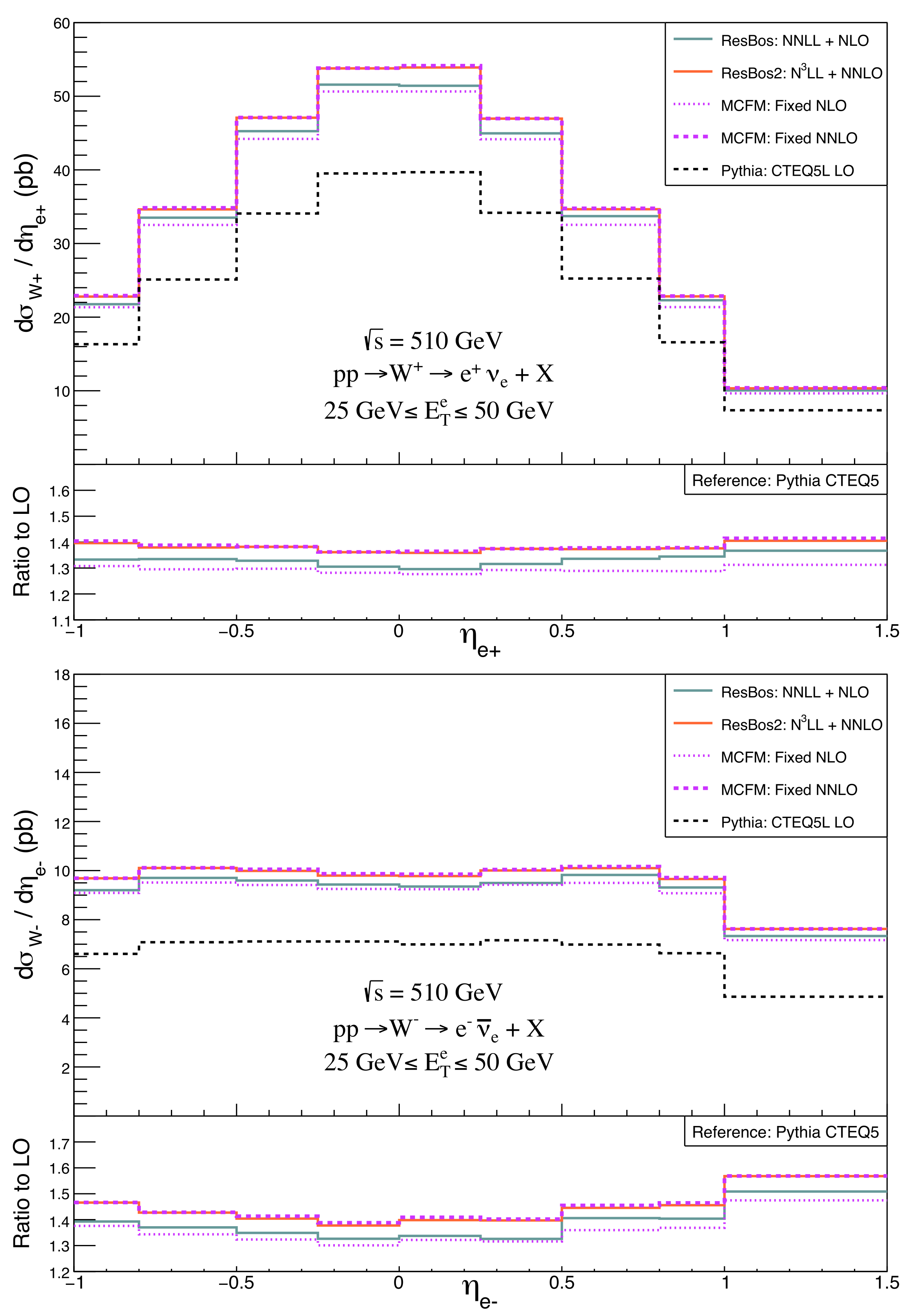}
\caption{Predicted lepton pseudorapidity differential cross sections (pb) for $W^+$ (upper) and $W^-$ (lower) boson production with lepton cuts at $\sqrt{s}=510$ GeV. Theoretical predictions are QCD-inclusive, \textit{i.e.} no cuts are imposed in the phase space of radiated gluons. $\mathcal{O}(\alpha_s^2)$ and $\mathcal{O}(\alpha_s)$ distributions differ in normalization by $\sim 7\%$.}
\label{fig:CS}
\end{figure}

NLO differential cross sections also receive radiative corrections from emission of real gluons and (anti) quarks, creating a spectrum of non-zero boson's transverse momentum $q_T$, shown in the right panel of Fig.~\ref{fig:dists}. From the perturbative expansion, the transverse momentum differential distribution is logarithmically enhanced at scales much smaller than the characteristic energy ($q_T \ll Q \sim M_W$), with the following behavior: 
\begin{equation}
    \frac{1}{\sigma} \frac{d\sigma}{dq_T^2} \approx \frac{1}{q_T^2}\left[ C_1 \alpha_s \ln \left( \frac{Q^2}{q_T^2} \right) + C_2 \alpha_s^2 \ln^3 \left( \frac{Q^2}{q_T^2} \right) + \ldots C_n \alpha_s^n \ln^{2n-1} \left( \frac{Q^2}{q_T^2} \right) \right],
\label{eq:FO}    
\end{equation}
where subleading logarithms and $(q_T/Q)$ power corrections are omitted. Consequently, missing higher-order terms contribute to the same extent as any fixed-order truncation, breaking down the convergence of the perturbative expansion. To restore convergence, large logarithms are resummed to all perturbative orders, leading to the Sudakov exponential factor, $e^{-\mathcal{S}}$ \cite{Collins:1984kg}
\begin{equation}
    \mathcal{S}(b,Q) = \int^{Q^2}_{(b_0/b)^2} \frac{d\mu^2}{\mu^2} \left[ \ln\left( \frac{Q^2}{\mu^2} \right) A \left(\alpha_s (\mu^2) \right) + B\left(\alpha_s (\mu^2) \right) \right],
\label{eq:Sud}
\end{equation}
with $\mu$ the resummation energy scale, $b$ the impact parameter, which is the Fourier conjugate of transverse momentum, and $A(\alpha_s)$, $B(\alpha_s)$ are perturbatively calculable coefficients. The Sudakov factor suppresses contributions from large logarithms, resulting in $d\sigma/dq_T \to 0$ as $q_T \to 0$ upon inverse Fourier transformation back to $q_T$ space with $b$-dependent parton distribution functions \footnote{Details of the resummation formalism for $W$ boson production at RHIC can be found in \cite{Nadolsky:2003fz,Nadolsky:2003ga} and for CSS resummation in \cite{Collins:1984kg,Campbell:2017hsr,Balazs:1997xd,Isaacson:2017hgb}.}. Since the resummed part is only valid at small $q_T$, to obtain a physical cross section, a component reflecting the convergent perturbative behavior at large $q_T$ and smoothly matched onto the resummed part is required, referred to as the $Y$ function, where the fixed-order expansion of the resummed part is subtracted from the overall fixed-order distribution to remove the singular behavior, giving negligible corrections at small $q_T$. The right panel of Fig. \ref{fig:dists} shows the $W^+$ transverse momentum differential cross section at fixed-order NLO with the log-enhanced behavior at small $q_T$, which is suppressed in the NNLL resummed + NLO matched distribution. In contrast, the two distributions have similar convergence for large $q_T$ values. In the subsequent comparisons, I also include resummed contributions from NNLO, contributing a few percent to the total rate and further modifying the $q_T$ dependence. The NNLL$+$NLO (\textsc{ResBos} \cite{Balazs:1997xd}) and N$^3$LL$+$NNLO (\textsc{ResBos2} \cite{Isaacson:2017hgb,Isaacson:2023iui,Isaacson:2022rts}) calculations expand $A$, $B$ and $Y$ functions up to $\alpha_s^3$, $\alpha_s^2$, $\alpha_s$ and $\alpha_s^4$, $\alpha_s^3$, $\alpha_s^2$ orders, respectively. Assessing the robustness of the experimental observable, Fig. \ref{fig:CS} shows $W^\pm$ inclusive differential cross sections in $e^\pm$ pseudorapidity at various fixed and resummed orders, with binning and $e^\pm$ fiducial cuts from the STAR experiment, \textit{i.e.} pseudorapidity within the asymmetric detector acceptance volume ($-1 < \eta_e < 1.5$) and transverse energy around the Jacobian peak ($25 \textrm{ GeV} \leq E_T^e \leq 50 \textrm{ GeV}$). These predictions are computed with the CT18 NNLO central PDF set \cite{Hou:2019efy}, setting the factorization and renormalization scales to the dynamical transverse mass $\mu = \sqrt{Q^2 + q_T^2}$. Both the resummed distributions and fixed-order ones are consistent, with minor differences within PDF and scale uncertainty, as expected, since rapidities are mostly sensitive to PDF shapes. NNLO distributions are relatively uniformly enhanced by $\sim7\%$ compared to the NLO ones.

\section{Examination of veto on QCD recoil with \textsc{Pythia}}
\label{sec:PYTHIA}
It is important to emphasize that, to be suitable for constraining the PDFs, $e^\pm$ pseudorapidity distributions like the one in the left panel of Fig.~\ref{fig:dists} and Fig.~\ref{fig:CS} must be fully inclusive in QCD radiation; this condition is the key assumption behind the provided factorization and resummation formulas. On the other hand, the STAR measurement vetoes hadronic activity inside the fiducial region in order to suppress QCD and electroweak backgrounds, thus breaking the inclusivity condition. Specific acceptance cuts and veto parameters in the STAR analysis can be found in Ref. \cite{nam2022measurementswzgammacross}. 

\begin{figure}[!ht]
\centering
\includegraphics[width=1\textwidth]{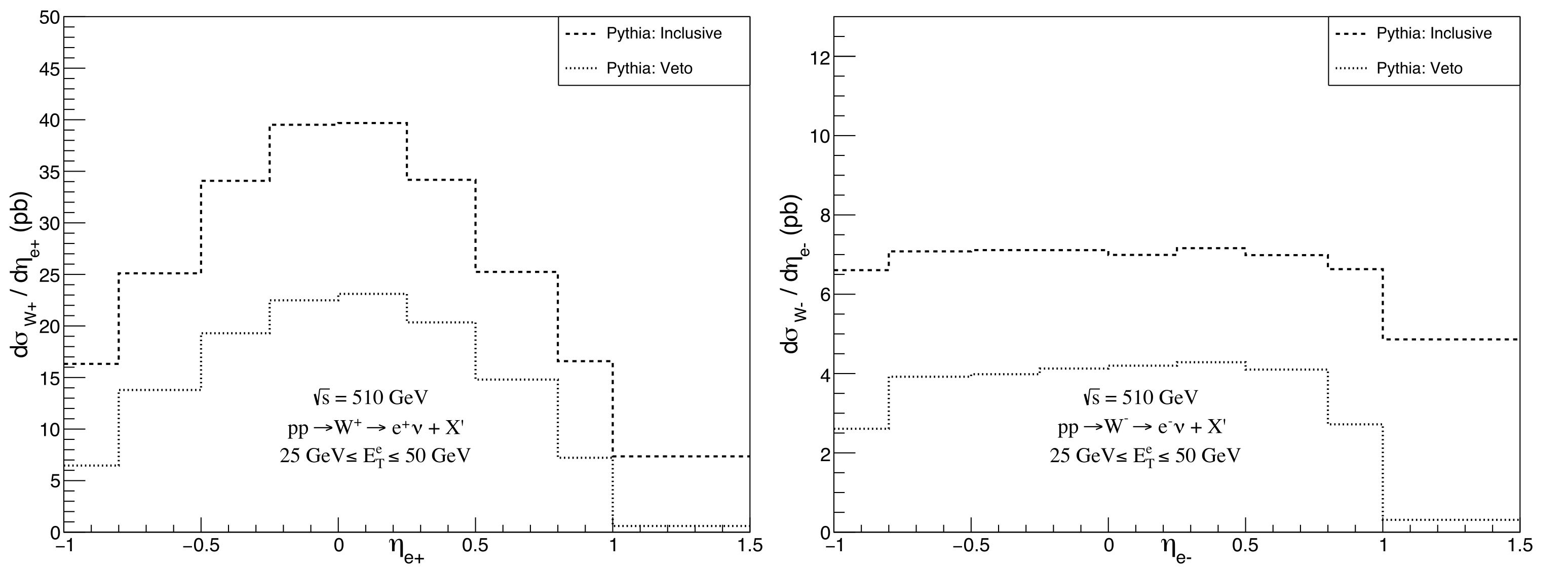}
\caption{Pythia LO lepton pseudorapidity cross sections (pb) for $W^+$ (left) and $W^-$ (right) cross sections at $\sqrt{s}=510$ GeV and using CTEQ5L PDFs, with and without vetoes on QCD radiation.}
\label{fig:pyth}
\end{figure}

To estimate the impact of the veto, the STAR collaboration employs leading-order plus parton shower Monte Carlo (MC) simulations. The impact of a veto is reviewed here, using one of the STAR simulations \cite{Nam:2025} with the \textsc{Pythia} parton shower \cite{Bierlich:2022pfr}. Comparing these predictions to the inclusive resummed ones poses a challenge at various levels, as the treatment of QCD radiation is not equal in the two approaches. First, NLO virtual corrections are not present in \textsc{Pythia}, i.e. the kinematically independent factor $K_{\textrm{NLO}} \sim (1+3.005\alpha_s) \approx 1.36$ is missing both in the \textsc{Pythia} total rate and differential distributions. Thus, the inclusive LO \textsc{Pythia} predictions shown in Fig.~\ref{fig:CS} are 30-40\% lower than the NLO ones. The impact of the missing virtual correction on the distribution shape is weak in the pseudorapidity range corresponding to the barrel calorimeter ($|\eta_e|<1$), as seen in the subpanels of Fig. \ref{fig:CS}, where the ratio of higher-order distributions to \textsc{Pythia} LO is taken. Note that the (N)NLO/\textsc{Pythia} ratio is slightly larger for $W^-$ in the STAR endcap calorimeter \cite{STAR:2002eml} pseudorapidity bin ($1<\eta_e<1.5$), possibly induced by the PDF choice in the \textsc{Pythia} simulation, since the STAR collaboration generates \textsc{Pythia} simulations with the leading-order CTEQ5L set \cite{Lai:1999wy}, which behaves substantially differently from the modern CT18 (N)NLO PDF sets. Second, a correction from the parton to detector level in \textsc{Pythia} independently reduces both $W^\pm$ cross sections by an additional factor of $\sim 40\%$, shown in Fig.~\ref{fig:pyth}. This behavior is approximately uniform, suggesting that the effects of the STAR experimental assumptions generally cancel out in the STAR barrel calorimeter region ($|\eta_e|<1$) \cite{STAR:2002ymp} once the $e^\pm$ pseudorapidity differential cross section ratio is obtained, as discussed in Section \ref{sec:Data}. In contrast, the differential cross section in the pseudorapidity bin ($1<\eta_e<1.5$) corresponding to the endcap calorimeter \cite{STAR:2002eml} is severely reduced by the vetoes.\\

\section{Comparison to STAR measurements of $W^+/W^-$ cross section ratios}
\label{sec:Data}
 
\begin{figure}[!ht]
\centering
\includegraphics[width=1\textwidth]{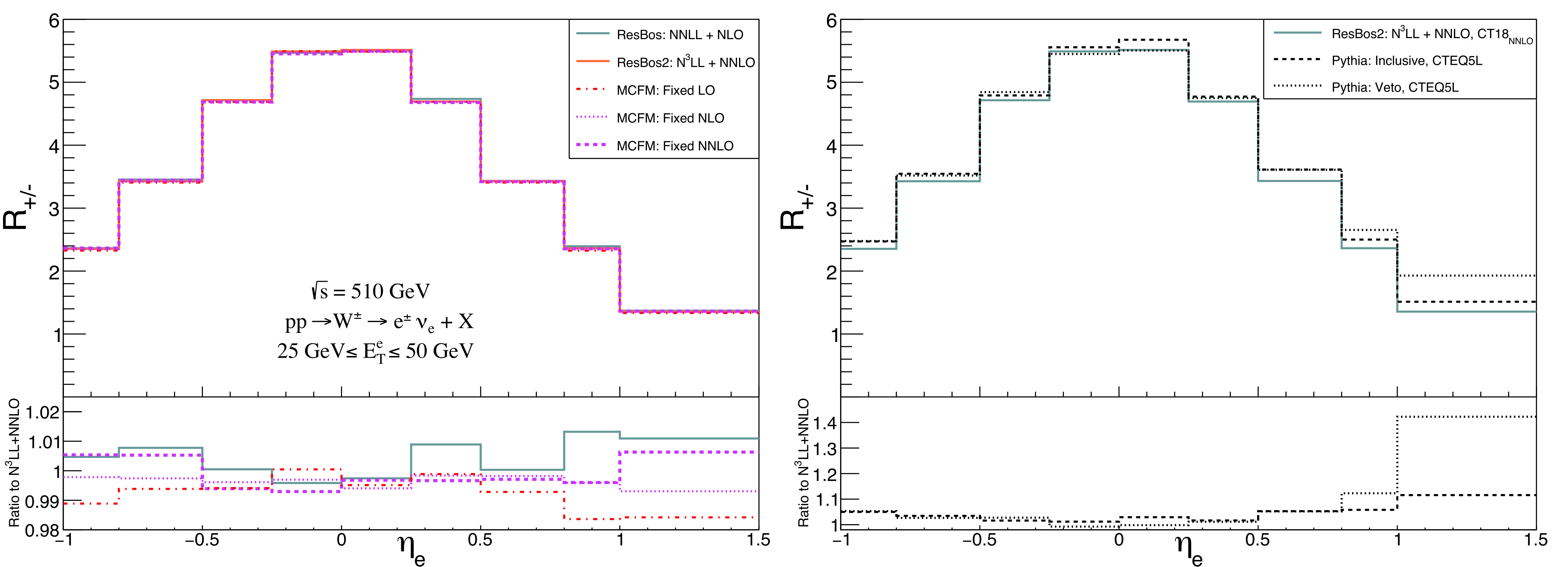}
\caption{Predictions for QCD-inclusive lepton pseudorapidity differential cross section ratio with various perturbative prescriptions for $W^\pm$ production at $\sqrt{s} = 510$ GeV.}
\label{fig:Ords}
\end{figure}

The left panel of Fig.~\ref{fig:Ords} shows the differential cross section ratio, $R_{+/-} \equiv (d\sigma_{W+}/d\eta_{e+})/(d\sigma_{W-}/d\eta_{e-})$, computed at the QCD-inclusive fixed-order LO and resummed NLO and NNLO with the CT18 NNLO PDF set. It is evident that the differences due to normalization and resummation in the $W^\pm$ cross sections from Fig.~\ref{fig:CS} largely cancel out in the QCD-inclusive cross section ratio. Comparing it at different orders and setting the N$^3$LL+NNLO distribution as the baseline, the lower subpanel of Fig.~\ref{fig:Ords} shows variations of at most $\sim 2\%$ across all pseudorapidity bins for distributions with the same PDF set, supporting the argument that this observable is robust with respect to perturbative corrections \cite{nam2022measurementswzgammacross}, while being informative about flavor asymmetry in the proton quark sea. The right panel of Fig.~\ref{fig:Ords} shows the cross section ratio from the \textsc{Pythia} simulations; the distributions with and without the hadronic veto are generally consistent, except at the bin of highest pseudorapidity where the effects of vetoes are substantial, noting that this region maps to the fractional momentum region where $\bar{d}(x)/\bar{u}(x)$ is in tension for deuteron scattering experiments. The inclusive N$^3$LL+NNLO prediction has a similar overall agreement with the veto distribution in the barrel calorimeter region, as shown in the lower subpanel where the ratio to the resummed distribution is taken. Deviations between the \textsc{ResBos2} and \textsc{Pythia} distributions also arise from the choice of a different PDF set, CTEQ5L \cite{Lai:1999wy}. Additionally, the ratio of \textsc{Pythia} distributions was obtained to estimate the scaling factors from vetoed to inclusive level. These scaling factors are employed in the PDF reweighting analysis in Section \ref{sec:Sens}.
\begin{table}[!ht]
\centering
\begin{tabular}{|c|c|c|c|c|c|}
\hline
\textbf{Pseudorapidity Bin} & $\abs{\eta_e} < 0.25$ & $0.25 < \abs{\eta_e} < 0.5$ & $0.5 < \abs{\eta_e} < 0.8$ & $0.8 < \abs{\eta_e} < 1.0$ & $1.0 < \eta_e < 1.5$ \\ \hline
\textbf{Scaling Factor} & 1.03 & 0.997 & 1.00 & 0.970 & 0.784 \\ \hline
\end{tabular}
\caption{Scaling factors to correct STAR data from veto to inclusive level, obtained from the ratio of \textsc{Pythia} inclusive to \textsc{Pythia} veto $R_{+/-}$ distributions, averaging over symmetric lepton pseudorapidity bins.}
\label{tab:factors}
\end{table}

\begin{figure}[!ht]
\centering
\includegraphics[width=0.9\textwidth]{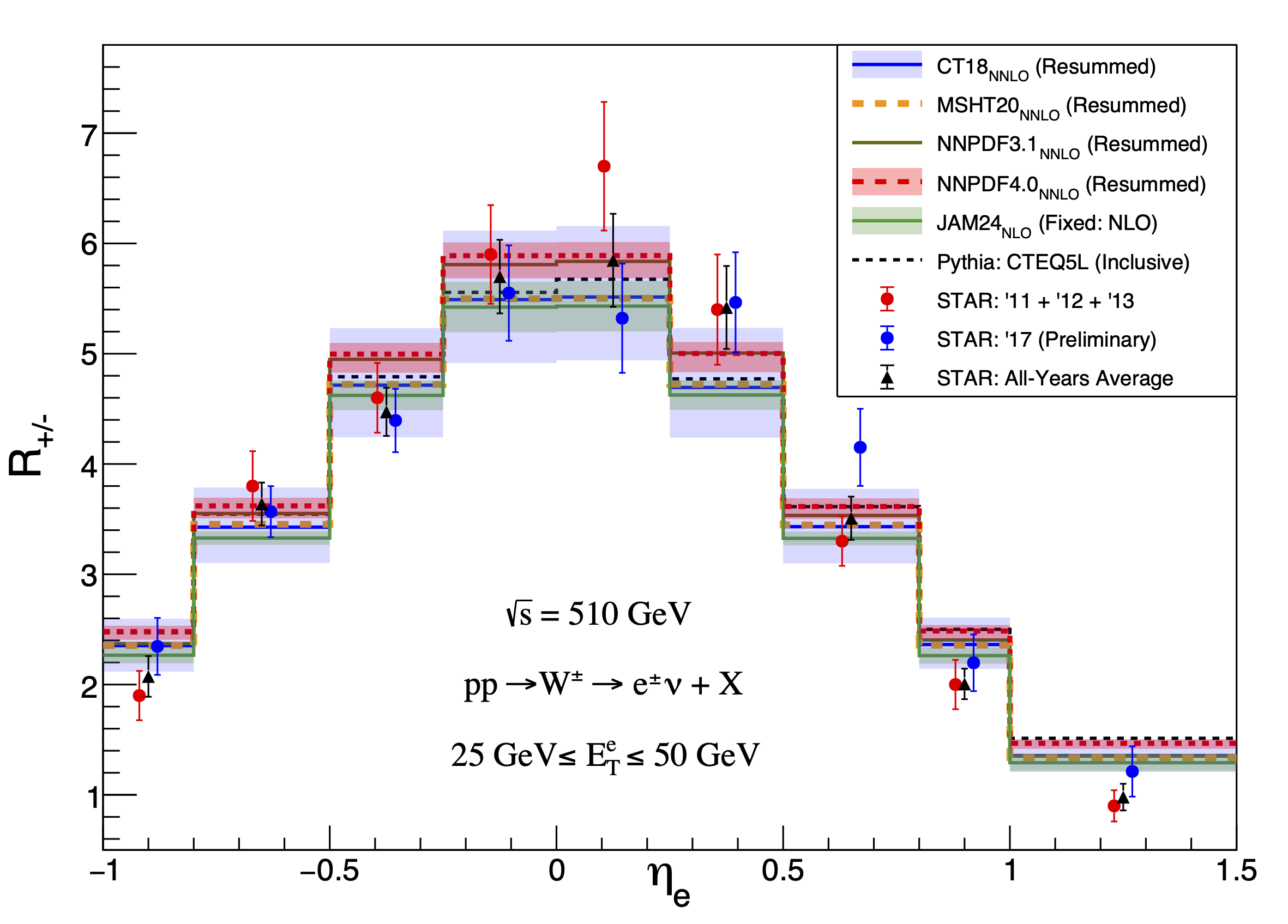}
\caption{Experimental and resummed $\sigma_{W+}/\sigma_{W-}$ cross section ratios for various PDF sets. Uncertainty bands are given at 68\% confidence level. Systematic and statistical uncertainties are combined in quadrature for all data points, and the veto to inclusive correction is not applied.} 
\label{fig:rat}
\end{figure}

Next, Fig.~\ref{fig:rat} examines the PDF variation of $R_{+/-}$, assuming that the perturbative prescription has negligible effects on the distribution. For \textsc{ResBos2} N$^3$LL+NNLO predictions, the CT18, MSHT20, NNPDF3.1 and NNPDF4.0 sets are considered, and JAM24 for fixed NLO with \textsc{MCFM-10.3}. Select PDF uncertainty bands are computed at 68 $\%$ confidence level for sets with different parameterization frameworks; CT18 for Hessian/polynomial, NNPDF4.0 for MC/neural network, and JAM24 for MC/polynomial. Both NNLO and NLO predictions are clustered in a central region mostly covered by the CT18 band, showing general agreement between the global fitting groups. Notably, JAM24
shows an improved description of STAR measurements compared to JAM19 \cite{Sato:2019yez}, due to the inclusion of NuSea and SeaQuest data in the fit \cite{Anderson:2024evk}.\\ 

\begin{table}[!ht]
\centering
\begin{tabular}{|c|c|c|c|c|c|c|}
\hline
\textbf{Set} & \textbf{CT18} & \textbf{MSHT20} & \textbf{NNPDF3.1} & \textbf{NNPDF4.0} & \textbf{JAM19} & \textbf{JAM24} \\ \hline
$\mathbf{\chi^2/N_{\textrm{\textbf{pt, Combined Pre-'13}}}}$ & 3.36 (2.88) & 3.24 (2.77) & 3.06 (2.63) & 4.26 (3.77) & 24.7 (24.1) & 2.89 (2.47) \\
$\mathbf{\chi^2/N_{\textrm{\textbf{pt, '17 (Preliminary)}}}}$ & 1.17 (1.08) & 1.10 (1.01) & 1.17 (1.15) & 1.44 (1.39) & 19.3 (19.6) & 1.31 (1.23) \\
$\mathbf{\chi^2/N_{\textrm{\textbf{pt, All-Years Average}}}}$ & 3.54 (2.88) & 3.35 (2.70) & 3.53 (2.96) & 5.21 (4.53) & 44.0 (43.2) & 2.90 (2.35) \\ \hline
\end{tabular}
\caption{Reduced $\chi^2$ for central predictions of various PDF sets to differential cross section ratio measurements from the STAR experiment with and without veto to inclusive correction factors. Since the bin of highest pseudorapidity is discarded for rescaling, its correction factor is arbitrarily set to 1 here. Accounting for correction in this bin significantly increases $\chi^2$ for all sets.}
\label{tab:chi2}
\end{table}

Table~\ref{tab:chi2} reports the reduced $\chi^2$ measure, assessing the agreement of central PDF predictions of $R_{+/-}$ with STAR's measurements from the preceding ('11+'12+'13, further referred to as \textit{combined pre-'13}) and latest ('17) datasets, as published (in parenthesis) as well as after approximately correcting the data to the inclusive level by scaling with factors from Table~{\ref{tab:factors}}, discarding the highest pseudorapidity bin due to the large differences with inclusive distributions. Note that the '17 dataset still shows a good agreement with the central NNLO PDF predictions after correction.\\

Consistently with an observation in Ref. \cite{STAR:2020vuq}, this study reports that the combined pre-'13 experimental points and inclusive theoretical distributions broadly follow the same trend. For this dataset, the $\chi^2$ is enhanced mainly due to the first, fifth, eighth, and ninth pseudorapidity bins, where the overlap between experimental error bars and central predictions with and without PDF uncertainties is marginal or null. Consequently, the reduced $\chi^2$ values in the first row of Table~\ref{tab:chi2} differ significantly from unity. In contrast, the '17 preliminary dataset yields significant improvements in $\chi^2$, as data points previously showing poor agreement generally lie near the cluster of central NNLO predictions, including the bin of highest pseudorapidity, where the theory-data overlap is now within uncertainty, thus shifting the reduced $\chi^2$ closer to unity. Finally, the average of all STAR datasets generally remains close to the NNLO predictions in most bins, however, $\chi^2$ does not reduce due to decrements in experimental error bars caused by higher statistics, particularly for the two last data points. 

\section{Estimated impact of STAR data on PDFs}
\label{sec:Sens}

\begin{figure}[!ht]
\centering
\includegraphics[width=1\textwidth]{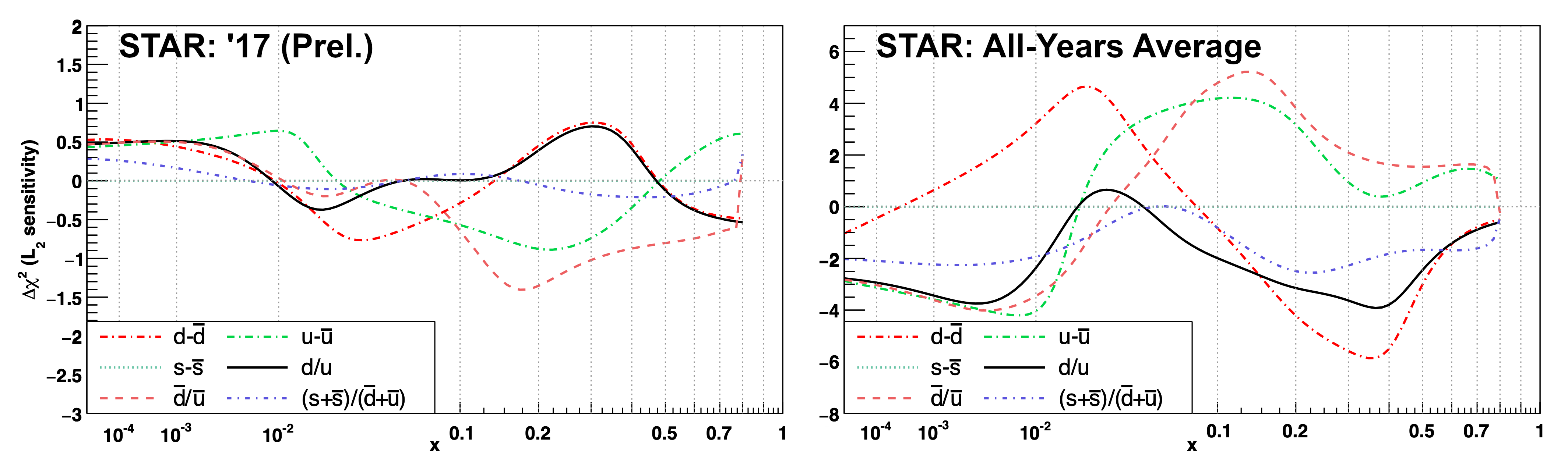}
\caption{CT18 NNLO sensitivity plots for PDF flavor combinations at $Q=2$ GeV, 68\% confidence level, with STAR's '17 preliminary (left) and all-years average (right) residuals.}
\label{fig:l2CT18}
\end{figure}

To evaluate the effect on sea-quark distributions prior to the inclusion of the latest STAR cross section ratio measurements in a PDF fit, sensitivity and reweighting analyses were performed on PDF sets published by the CTEQ-TEA group. First, following the studies in Refs. \cite{Jing:2023isu,Hobbs:2019gob,Guzzi:2021fre}, the statistical pull induced on PDFs by the data-theory residuals is analyzed through the $L_2$ sensitivity:
\begin{equation}
    S_{f,L_2}^H(x,Q) = \delta_H (\chi^2)\ C_H(f(x,Q), \chi^2),
\label{eq:sensit}
\end{equation}
where $\delta_H (\chi^2)$ is the Hessian symmetric uncertainty of STAR's total $\chi^2$ and $C_H(f(x,Q), \chi^2)$ is the Pearson correlation cosine between a PDF $f(x,Q)$ and $\chi^2$. From Ref. \cite{Hou:2016sho}, for a variable $X$ depending on a PDF fitted with $D$ parameters, the Hessian uncertainty at the confidence level determined by the PDF set is defined as:
\begin{equation}
    \delta_H(X) = \frac{1}{2}\sqrt{\sum_{i=1}^D \left[ X_{+i} - X_{-i} \right]^2},
\label{eq:delta}
\end{equation}    
with $X_{\pm i}$ denoting the variable evaluated with the $i$-th positive/negative eigenvector error PDF, while the correlation cosine is given by:
\begin{equation}
    C_H(f(x,Q), \chi^2) = \frac{1}{4\delta_H (f(x,Q))\  \delta_H (\chi^2)} \sum_{i=1}^D \left[ f_{+i}(x,Q) - f_{-i}(x,Q) \right] \left[ \chi^2_{+i} - \chi^2_{-i} \right].
\label{eq:cos}
\end{equation}
Therefore, $S_{f,L_2}^H(x,Q)$ estimates the change in $\chi^2$ when the PDF increases by one error unit, at given $x$ and $Q$ values, with negative sensitivity values implying an anticorrelation between $\chi^2$ and PDF increments, consequently pulling $f(x,Q)$ upward to better match the experimental observable. Conversely, positive sensitivities indicate downward pulls on $f(x,Q)$. For this analysis, the CT18 central set and its 58 error eigenvectors ($D=29$ in Eqs. \ref{eq:delta},\ref{eq:cos}) are chosen. The residuals are computed by comparing the $R_{+/-}$ '17 preliminary and all-years averaged experimental measurements with the NNLL+NLO predictions from all CT18 error sets. The $L_2$ sensitivity to residuals with the '17 dataset is shown in the left panel of Fig. \ref{fig:l2CT18} for indicated combinations of CT18 NNLO PDFs at $Q=2$ GeV. Specifically, $\bar{d}(x)/\bar{u}(x)$ shows a negative sensitivity, with the reduction of $\chi^2$ peaking near $x \sim 0.2$, precisely covered by the detector fiducial volume and indicating that the STAR '17 measurements prefer a slightly enhanced ratio in the large-$x$ region. Examination of the sensitivities of individual PDF flavors reveals that this pull is primarily induced by a reduction in $\bar{u}(x)$, associated with a positive value of the corresponding sensitivity \footnote{Sensitivity plots for individual PDF flavors are omitted for brevity.}. In contrast, the combined dataset (right panel) drastically generates the opposite effect with stronger downward $\chi^2$ pulls on the $\bar d/\bar u$ ratio induced by a significant increase of $\bar{u}(x)$ at $x \sim 0.1$.\\

The most recent global QCD analysis by the CTEQ-TEA collaboration produced CT25 \cite{Ablat:2025gdb,Ablat:2026CT25}, a NNLO PDF set with newly incorporated LHC data and alternative prescriptions to account for epistemic uncertainty, \textit{i.e.} pertaining to PDF parameterizations and fitting methodology. Among those, the CT25 flat prior NNLO PDF set (further referred to as \textit{CT25FlatP}) is of particular interest for this work. CT25FlatP, fully described in the upcoming journal publication \cite{Ablat:2026CT25}, is made up of $N=350$ PDF central solutions describing the proton-only subset from the global fit's data pool \textit{as good} as the nominal CT25 NNLO set. The distinguishing feature of CT25FlatP is that all of its sets are fits with good $\chi^2$ for the proton-only subset of the CT25 baseline, obtained either by statistical fluctuations of the data points or by changes in the PDF parametrizations. The distribution of the CT25FlatP sets can thus be taken as a prior that can be updated by the STAR measurements through reweighting, offering yet more constraints to the proton sea-quark PDFs with the advantage of being free of potential bias from nuclear effects. Before reweighting, the density of the CT25FlatP predictions does not carry a probabilistic meaning, as all 350 input sets are equally acceptable in this approach. The CT25FlatP prior uncertainty is therefore given by an envelope of all predictions. After reweighting, the posterior probability of a CT25FlatP member set is given by a generalization of the Giele-Keller weights described in \cite{Giele:1998gw}, which depend on the $\chi^2$ of the experimental measurement. Explicit formulas for reweighting are given in the Appendix.

\begin{figure}[!ht]
\centering
\includegraphics[width=1\textwidth]{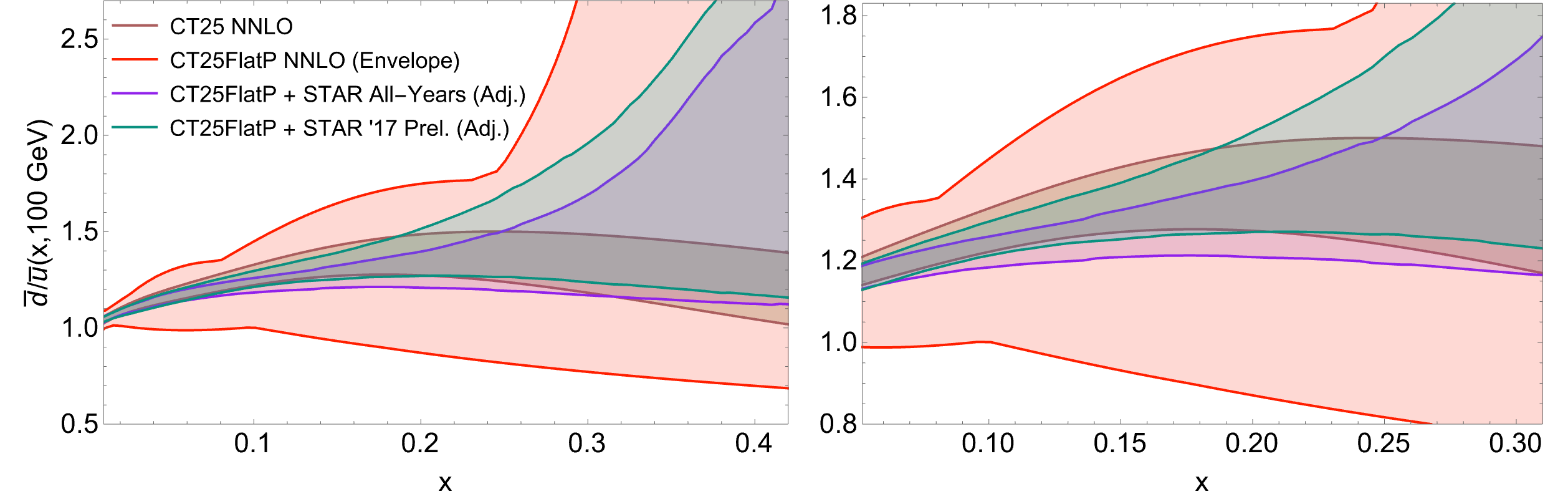}
\caption{$\bar{d}(x)/\bar{u}(x)$ uncertainty bands at $Q=100$ GeV for the CT25 NNLO and CT25FlatP sets. Reweighted bands are computed asymmetrically with adjusted weights from Eq.~(\ref{eq:weightAdj}). The right figure is an amplification of the left figure into the highest sensitivity fractional momentum region from the STAR experiment.}
\label{fig:dens}
\end{figure}

Motivated by the constraining power of the STAR measurements, Fig.~\ref{fig:dens} shows the CT25FlatP uncertainty for $\bar d(x,Q)/\bar u(x,Q)$ at $Q=100 $ GeV before and after reweighting by the $R_{+/-}$ ratio from the '17 preliminary and all-years averaged datasets. The STAR data is corrected to the QCD-inclusive level using the factors in Table~\ref{tab:factors}, with the highest pseudorapidity bin ($1 < \eta_e < 1.5$) excluded from this analysis due to the large difference between the vetoed and inclusive $R_{+/-}$ predictions, as well as its large $\chi^2$. The error bands from the nominal CT25 NNLO at 68\% confidence level (in brown) and the ``flat prior'' CT25FlatP (in red) ensembles are compared first. In the former, the $\bar{d}(x)/\bar{u}(x)$ ratio is constrained at $x>0.1$ by a combination of $\sigma_{pd}/2\sigma_{pp}$ measurements by the NuSea \cite{NuSea:2001idv} and SeaQuest \cite{SeaQuest:2021zxb,FNALE906SeaQuest:2025kjo} fixed-target Drell-Yan experiments. This constraint is not imposed on the CT25FlatP set, so the $\bar{d}(x)/\bar{u}(x)$ uncertainty of this set is much wider than the CT25 one. Large variations in this region are reflected by the envelope of CT25FlatP solutions, realized for some parameterizations.\\
 
Fig.~\ref{fig:dens} also shows $\bar{d}(x)/\bar{u}(x)$ uncertainty bands obtained by reweighting with the '17 preliminary and all-years averaged STAR datasets (green and violet, respectively), with weights corresponding to the $\Delta\chi^2=1$ tolerance criterion for the PDF uncertainty, suitable when the given experiment dominates without tension with other experiments, and given that the CT25FlatP prior already accounts for the PDF parametrization uncertainty. \footnote{This uncertainty may be increased in the full global analysis if there are tensions with other experiments.}. The reweighted PDFs are consistent with CT25 NNLO, although the all-years averaged dataset, with its elevated overall $\chi^2$ for CT18 NNLO in Table~\ref{tab:chi2}, prefers a lower $\bar{d}(x)/\bar{u}(x)$ in the most sensitive $x$ region. Overall, the STAR measurement is compatible with $\bar d/\bar u \approx 1.25$ at $x\approx 0.15$ already expected based on the NuSea/SeaQuest observations. It is substantially incompatible with $\bar{d}(x)/\bar{u}(x)$ equal to one or less than one, values that might be allowed in other analyses \cite{Ma:2025aga,Ruzi:2026igh}.\\

\section{Conclusions}
\label{sec:Conc}

In this work, I have explored the impact of hadronic vetoes on $W^\pm$ boson production $e^\pm$ pseudorapidity differential cross sections at RHIC, focusing on their ratio, which is sensitive to the sea-quark flavor asymmetry in the proton. Comparing fixed-order and resummed inclusive predictions with leading-order parton showering simulations, I observed that differences arising from normalization and the hadronic recoil veto mostly cancel out in the cross section ratio for pseudorapidities within the STAR barrel calorimeter volume, although notable deviations induced by the veto remain in the endcap calorimeter region, with Table~{\ref{tab:factors}} listing the approximate factors to correct the $\sigma_{W^+}/\sigma_{W^-}$ ratios from the vetoed to QCD-inclusive level. Then, the preliminary dataset of STAR's pseudorapidity ratio measurements in 2017 was compared with inclusive resummed central predictions using various modern PDF sets, showing notable reductions in $\chi^2$ in contrast to the STAR measurements from the earlier years. Both an $L_2$ sensitivity analysis done with the CT18 NNLO PDF set and reweighting of the CT25 Flat Prior set indicate that the STAR result agrees with $\bar d(x,Q)/\bar{u}(x,Q)\approx 1.25$ at $x=0.15$ and $Q=100$ GeV. Reweighting the CT25FlatP \textit{proton-only} set with the STAR data significantly reduces the uncertainty on $\bar{d}(x)/\bar{u}(x)$, see Fig.~\ref{fig:dens}. The reweighted $\bar{d}(x)/\bar{u}(x)$ ratio at $x \approx 0.15$ is slightly lower, specially for the combined STAR dataset from all years, than the nominal CT25 one which includes the Nusea/SeaQuest $\sigma_{pd}/(2\sigma_{pp})$ Drell-Yan data, but remains in agreement with CT25 within the reweighted PDF uncertainty while being inconsistent with the SU(2)-symmetric $\bar d/\bar u=1$. Therefore, this work shows that the $\sigma_{W^+}/\sigma_{W^-}$ cross section ratio measurements by the STAR collaboration can constrain the light sea-quark distributions in the proton with the advantage of being free of nuclear effects. The main findings presented here are based on the preliminary 2017 STAR measurements and on the application of PDF sensitivity and reweighting techniques to published PDF sets. A more complete assessment will be obtained after the publication of the final STAR results by including them into the CTEQ-TEA global fit, together with other sensitive fixed-target deuteron and LHC measurements.

\section{Acknowledgments}
I would like to thank Pavel Nadolsky, C.-P. Yuan, and colleagues at Michigan State University for insightful discussions and detailed instruction on the usage of the resummation codes. Also, I want to thank Jae D. Nam for the close communication regarding the STAR experiment, and Aurore Courtoy as well as her research group for fruitful interactions at IFUNAM. This work was supported by the U.S. National Science Foundation under Grant No. PHY-2310291; by the Wu-Ki Tung Endowement and partially by the UNAM Grant No. DGAPA-PAPIIT IN102225.

\appendix
\section{Reweighting}
The CT25FlatP PDF set can be updated with information from new experimental data thanks to Bayes theorem \footnote{Further details and explicit derivations can be found in \cite{Giele:1998gw} and \cite{Ball:2010gb}.}; the posterior distribution is given by the normalized product of the prior distribution and the likelihood of describing the new data, which depends on the $\chi^2$ measure. To each member set corresponds a weight,
\begin{equation}
    w_k \equiv \frac{a_k\ e^{-\frac{1}{2} \chi_k^2}}{\frac{1}{N} \sum_{k=1}^{N} a_k\ e^{-\frac{1}{2} \chi_k^2}},
\label{eq:weightAdj}    
\end{equation}
where the sum of weights is normalized to the number of members of the PDF set, $\sum_{k=1}^N w_k =N$. For this analysis, $\chi^2_k$ is constructed from STAR central values and uncertainties for $R_{+/-}$ corrected to the QCD-inclusive level, not including the highest $\eta_e$ bin, as well as from the NNLL+NLO prediction for $R_{+/-}$ based on the $k$-th CT25FlatP member set.
Then, the expectation value and uncertainty of a PDF-dependent quantity $X$ are estimated by averaging over a finite number of member sets, whose importance is assessed not only through their likelihood, but also for the mechanism through which they were sampled, accounted for by the factor $a_k$ in Eq. (\ref{eq:weightAdj}). Therefore, the central (expectation) value of $X$ is given by the weighted average over $N=350$ CT25FlatP member sets:
\begin{equation}
    \mu_{X} \equiv \frac{1}{\sum_{k=1}^N w_k} \sum_{k=1}^N w_k X_k,
\label{eq:weightMean}
\end{equation}
and the reweighted positive and negative uncertainties are asymmetric weighted standard deviations with Bessel correction, computed by
\begin{equation}
\begin{split}
    \delta^+X = + \sqrt{\frac{\sum_{X_k>\mu_{X}} w_k}{\left(\sum_{X_k>\mu_{X}} w_k\right)^2 - \sum_{X_k>\mu_{X}} w^2_k } \cdot \sum_{X_k>\mu_{X}} w_k \cdot \left(X_k-\mu_{X} \right)^2 },\\
    \delta^-X = -\sqrt{\frac{\sum_{X_k<\mu_{X}} w_k}{\left(\sum_{X_k<\mu_{X}} w_k\right)^2 - \sum_{X_k<\mu_{X}} w^2_k } \cdot \sum_{X_k<\mu_{X}} w_k \cdot \left(\mu_{X}-X_k \right)^2 }.
\end{split}    
\label{eq:Uncertainties}
\end{equation}

The choice of $a_k=1$ in Eq. (\ref{eq:weightAdj}) corresponds to the original Giele-Keller weight \cite{Giele:1998gw} with equal sampling importance, introduced to update probabilities of discrete member sets outside of full refitting only through the likelihood. As the members of the prior CT25FlatP set are not distributed uniformly, in general, non-unit weight factors ($a_k \neq 1$) are considered to improve the convergence of posterior estimates. For example, $a_k$ can be constructed using a conformal coverage technique.

\begin{figure}[!ht]
\centering
\includegraphics[width=0.6\textwidth]{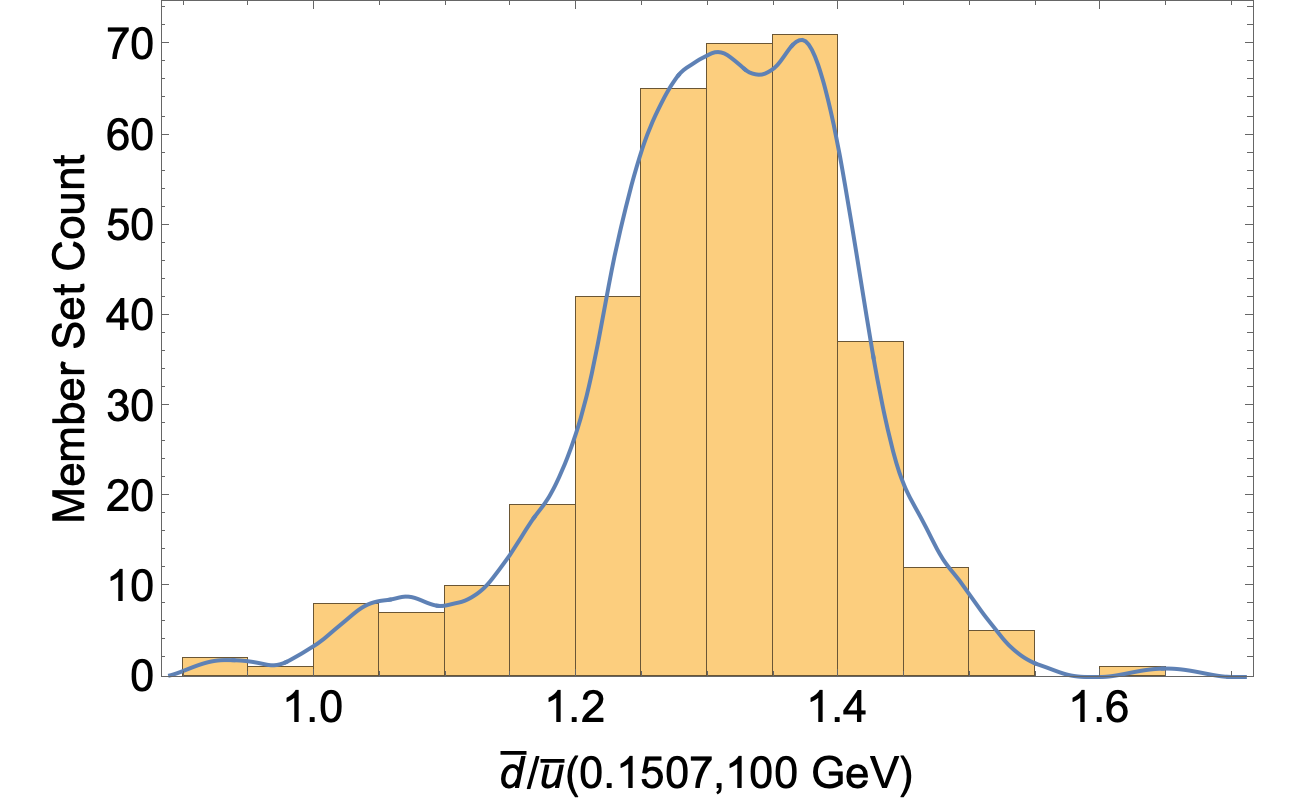}
\caption{Distribution of $\bar{d}(x)/\bar{u}(x)$ values spanned by the PDF member sets of CT25FlatP prior to reweighting at $x=0.1507$ and $Q=100$ GeV. The solid line corresponds to a biweight kernel density estimation.}
\label{fig:KDE}
\end{figure}

Since the CT25FlatP ratio $\bar d/\bar u$ at $x\approx 0.15$ and $\approx 100 $ GeV, corresponding to $y_W \approx 0$ within the STAR kinematic coverage, is only weakly correlated with other PDF flavors, the optimal coefficients $a_k$ can be estimated directly from the $\bar{d}/\bar{u}$ distribution of the CT25FlatP member sets. Fig. \ref{fig:KDE} shows that the $\bar{d}/\bar{u}$ distribution is more densely sampled in the $[1.2,1.4]$ interval due to the way CT25FlatP was constructed, in particular, because 200 member sets originate from two fits including deuteron data. Nonetheless, the less common member sets at the edges of the span (\textit{e.g.} obtained with distinct PDF parameterizations that are allowed in the absence of deuteron constraints) are as good as the ones in the middle region, given their nearly equal $\chi^2$ for the proton-only baseline of CT25. Accounting for the non-uniform distribution of this quantity, the weights in Eq. (\ref{eq:weightAdj}) are adjusted to yield Horvitz-Thomson estimators \cite{Horvitz01121952}, \textit{i.e.} with $a_k$ factors inversely proportional to the $\bar{d}/\bar{u}$ prior density of member sets, which is found through kernel density estimation (KDE), shown in Fig. \ref{fig:KDE}. From the observed distribution, the $a_k$ factors increase $w_k$ for member sets at the tails, while lowering the weights within the central $[1.2,1.4]$ interval. Thus, the mean and standard deviation for $\bar{d}/\bar{u}$ from Eqs. (\ref{eq:weightMean}) and (\ref{eq:Uncertainties}) account for the density of prior member sets under this approach. It turns out that either reweighting approach, with the original Giele-Keller weights ($a_k=1$) or with the KDE adjustment, yield very similar estimates; member sets with the highest likelihood to the new data generate approximately constant $a_k$ factors in the $[1.2,1.4]$ interval where the bulk of the distribution lies, effectively reproducing the standard reweighting Giele-Keller method in this region. 

Moreover, the consistency of the prior distribution with the newly included STAR '17 dataset is partially assessed by the number of effective replicas remaining after reweighting, given by $N_\textrm{Eff} = \exp(\frac{1}{N} \sum_{k=1}^N w_k \ln(N/w_k))$ \cite{Ball:2010gb}.
This number is approximately the same under both prescriptions; 186 with the original Giele-Keller weights and 191 with the adjusted weights, therefore, the STAR '17 dataset provides moderate constraints to the proton sea-quark distributions. Due to this observation, only the prescription of adjusted weights was considered in Section \ref{sec:Sens}.\\

\bibliographystyle{utphys} 
\bibliography{main}

\end{document}